\documentclass[aps,prb,superscriptaddress,reprint]{revtex4-1}%
\usepackage{epsfig,pslatex,latexsym,times,amssymb,amsmath,graphicx}
\usepackage{bm, float, lipsum}
\usepackage{amsmath}
\usepackage{amsfonts}
\usepackage{amssymb}
\usepackage{mathrsfs}
\usepackage{color}
\usepackage{graphicx}%
\providecommand{\U}[1]{\protect\rule{.1in}{.1in}}
\begin{document}
\author{N. J. Harmon}
\email{nicholas-harmon@uiowa.edu} 
\author{M. E. Flatt\'e}
\affiliation{Department of Physics and Astronomy and Optical Science and Technology Center, University of Iowa, Iowa City, Iowa
52242, USA}
\date{\today}
\title{The effects of spin-spin interactions on magnetoresistance in disordered organic semiconductors}
\begin{abstract}
A recent theory of magnetoresistance in positionally disordered organic semiconductors is extended to include exchange and dipolar couplings between polarons.
Analytic results are discovered when the hyperfine, exchange, and dipolar interactions have little time to operate between hopping events.
We find an angle-of-field dependence of the magnetoresistance that agrees with previous experiments and numerical simulations. 
In addition we report new magnetoresistive behavior that critically depends upon the amount of anisotropy in the dipolar interaction.
\end{abstract}
\maketitle

\section{Introduction}

Some spintronic applications are hindered in many inorganic semiconductors due to the heavy constituent atoms present and the resulting strong spin-orbit coupling that severely limits spin lifetimes.
The  less massive constituent atoms of some organic systems offer the advantage of weak spin-orbit coupling as well as other benefits such as cheap processing and chemical tuning.\cite{Naber2007, Vardeny2010, Bergenti2011}
In recent years, spurred by magnetoresistance measurements of 10-20\% in small fields at room temperature,  the effects of magnetic fields on currents through organic layers have been studied by many researchers with hopes of adding magnetic functionality to existing and new organic devices.\cite{Kalinowski2003,Francis2004,Prigodin2006,Desai2007,Hu2007,Bloom2007,Bobbert2007,Bergeson2008,Wagemans2010, Nguyen2010} 
This large magnetoresistive phenomenon, known as organic magnetoresistance (OMAR), has typically been explained in terms of spin correlations between charge carriers or polaron pairs. 
There is strong evidence\cite{Nguyen2010b} that OMAR occurs whether the spin correlations are between like charge carriers\cite{Bobbert2007, Harmon2012a, Harmon2012b} or oppositely charged carriers.\cite{Prigodin2006, Desai2007, Hu2007}
By taking note of the intimate relationship between spin and charge transport,\cite{Flatte2000b} two recent works\cite{Harmon2012a, Harmon2012b} have proposed a theoretical description of OMAR based on percolation theory in positionally disordered unipolar organic systems.
The goal of the current article is to apply this theory to include exchange and dipolar interactions between polaron pairs and thereby establish the wider applicability of the theory.

Wagemans \emph{et al.} studied the problem of incorporating spin-spin interactions experimentally, by measuring the field angle dependence of OMAR, and theoretically, by  using the stochastic Liouville equation approach.\cite{Wagemans2011}
Prior to the measurements of Wagemans \emph{et al.} no field angle dependence in OMAR had been reported, though in 1970's magnetic field angular dependence had been observed in the photoluminescence of an organic crystal.\cite{Groff1974}
Wagemans \emph{et al.}  also posit that anisotropic hyperfine interactions can lead to similar angular dependence in OMAR behavior. 
Anisotropic hyperfine interactions are not considered  in this  article.

We apply spin-dependent percolation theory to organic magnetoresistance in the presence of exchange and dipolar interactions and (1) obtain analytic expressions for the angular dependence of the OMAR, including the number of extrema as a function of angle, in the regime of fast hopping, (2) provide a heuristic picture of the effect of exchange on OMAR, by inferring that it behaves as an effective hopping for spin pairs, and finding an analytic correspondence between hopping and exchange valid for all regimes of hopping, and (3) predict large negative MR for special angles where the spin splitting from exchange and dipolar interactions vanishes in the high-field regime. 
In the regime where analytic expressions can be obtained, we find in the limit of hopping times going to zero that the angular dependence vanishes (in agreement with the numerical results of Ref. \onlinecite{Wagemans2011}). 
This theory also explains, via analytic expressions, how two unique angular dependences of the saturated magnetoresistance develop for certain ratios of the exchange and dipolar constants, and describes the interplay between them when both are present;
in all three scenarios magnetoresistive behavior distinct from the non-interacting case is seen. 
Notably, spin-spin interactions allow for the presence of an ultra-small-field-effect (USFE), where the magnetoresistance changes sign in small fields.
We further find that the amount of anisotropy in the dipolar displacement vector is crucial in determining the angle dependence of the magnetoresistance. 

Section \ref{section:semiclassical} describes the semiclassical treatment of the hyperfine fields along with the exchange and dipolar spin-spin interactions, and for completeness of this article Section \ref{section:model} summarizes the percolation-based model of OMAR from Refs. \onlinecite{Harmon2012a} and \onlinecite{Harmon2012b}.  This model is then applied in Section~\ref{section:spinspin} to the fast hopping regime of carrier transport in organics in the presence of exchange and dipolar spin-spin interactions, and analytic results   are derived.
Due to the presence of spin-spin interactions, the  theory outlined in Ref. \onlinecite{Harmon2012b} for OMAR in a regime of arbitrary hopping times must be extended; 
section \ref{section:general} introduces a different formalism that can include spin-spin interactions and arbitrary hopping rates, and then describes the full magnetoresistance problem with arbitrary hopping rates and inclusion of exchange and dipolar spin-spin interactions.

\section{The hyperfine, exchange, and dipolar interactions}\label{section:semiclassical}

Consider a polaron pair (PP); the physical picture is two spin-$\frac{1}{2}$ polarons located at two sites. 
The spins evolve coherently under the influence of identical applied fields, different classical nuclear or hyperfine fields, \emph{and} the exchange and dipolar fields. 
The respective PP Hamiltonian is
\begin{equation}\label{eq:hamiltonian}
\mathscr{H} = \mathscr{H}_{Z} +\mathscr{H}_{hf} +\mathscr{H}_{exc} + \mathscr{H}_{dip} 
\end{equation}
whose terms are expressed later in this section.
All Hamiltonians and spin operators are written in units of $\hbar$ throughout.
Since nuclear dynamics are slow, the total hyperfine field at a site is stationary and composed of many nuclei (Figure \ref{fig:hyperfineDiagram}), which justifies the classical approximation.\cite{Rodgers2007}
The hyperfine fields are random site-to-site. 
When a polaron hops to an unoccupied site (disassociation), its coherent spin evolution ceases as the polaron now feels a new local magnetic field. When the polarons hop to combine as a bipolaron, they are necessarily in the singlet state and the large exchange interaction prevents further spin evolution. In the PP, the two polarons are further apart, exchange is much smaller which allows spin transitions. If the hopping is fast, the PP spin evolves very little due to the motional narrowing effect. When the spins reside on their sites for a long period of time, they precess around their hyperfine fields many times and are ``fully mixed".

\subsection{The hyperfine interaction}

The Larmor frequency due to the conglomerate nuclear spin is the constant classical vector
\begin{equation}
\bm{I}_{N_i} = \sum_j a_j \bm{I}_j
\end{equation}
where $a_j$ is the hyperfine coupling constant between the electron and the $j$-th nucleus.
$\bm{I}_{N_i}$, the precession rate for carrier at site $i$, is made up of many different nuclei each with vector length $a_j \sqrt{I_j(I_j+1)}$ pointing in a random direction. $I_j$ is a spin quantum number (1/2 for a proton). 
The probability distribution for finding an $i$ `molecule' among an ensemble of such molecules with its total end-to-end vector between $\bm{I}_{N_i}+d\bm{I}_{N_i}$ is \cite{Flory1969,Schulten1978}
\begin{equation}\label{eq:flory}
W(\bm{I}_{N_i}) = \Big( \frac{2}{ \pi a_{eff}^2} \Big)^{3/2} \exp(-2 \frac{I^2_{N_i}}{a_{eff}^2})
\end{equation}
where
\begin{equation}
a^2_{eff} = \frac{4}{3} \sum_j a_j^2 I_j(I_j+1) .
\end{equation}
is the effective hyperfine coupling width due to all the nuclei at a site.
We consider all sites to be of the same type.  The effective \emph{field} is $B_{eff} = a_{eff}/\gamma_e$ with $\gamma_e = 0.176$ ns$^{-1}$ mT$^{-1}$. 
$B_{eff}$ is on the order of 1 mT.

\begin{figure}[ptbh]
 \begin{centering}
        \includegraphics[scale = 0.45,trim = 210 250 80 200, angle = -0,clip]{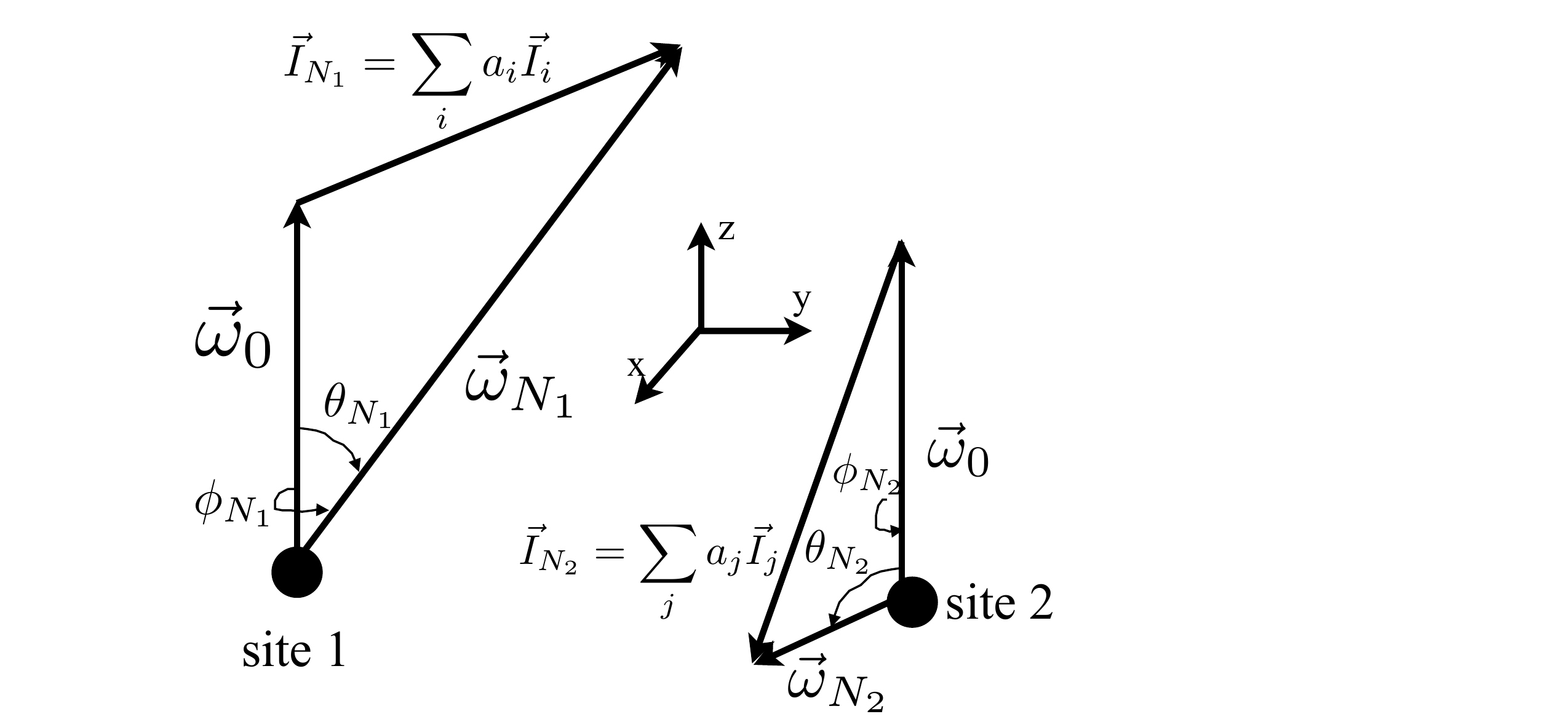}
        \caption[]
{Semiclassical description of total field. Each site is composed of some number of spin-moment-carrying nuclei which combine with the external field $\bm{\omega}_0$ to create $\bm{\omega}_N$. Each site has a different total field picked from the distribution of Eq. (\ref{eq:flory}).}\label{fig:hyperfineDiagram}
        \end{centering}
\end{figure}

A polaron spin precesses at a frequency
\begin{equation}
\bm{\omega}_{N_i} = \bm{I}_{N_i}  + \bm{\omega}_0,
\end{equation}
where $\bm{\omega}_0 = \gamma_e B\hat{z}$ is the applied field.
When neglecting spin-spin interactions, the non-interacting Hamiltonian of interest is
\begin{equation}
\mathscr{H}_0 = \mathscr{H}_{hf} + \mathscr{H}_{Z} 
 = 
 (\bm{I}_{N_1} + \bm{\omega}_0) \cdot \bm{S}_1 +  (\bm{I}_{N_2} + \bm{\omega}_0) \cdot \bm{S}_2
 \end{equation}
whose terms can be identified in Figure \ref{fig:hyperfineDiagram}.

In Section \ref{section:model} we discuss quantities that require hyperfine averaging.
Explicit averaging by integrating over the six hyperfine variables is a time-consuming numerical task. 
We opt to average by considering many different hyperfine configurations taken from the Gaussian distribution. 
Typically we use $10^3 - 10^4$ configurations.

\subsection{The exchange and dipolar interactions}

Due to the Coulomb interaction and the quantum mechanical requirement of an antisymmetric total wavefunction, 
two polarons experience a mutual exchange interaction:
\begin{equation}
\mathscr{H}_{exc} = J(\bm{r}) (\frac{1}{2} + 2 \bm{S}_1 \cdot \bm{S}_2),
\end{equation}
where $2 J(\bm{R}) $ is the exchange coupling constant which defines the T-S energy splitting (in frequency units). $J(\bm{R}) $ decreases exponentially with distance between the spin pair.
Since PPs have a large assortment of separations ($R$), the exchange interaction is expected to vary widely throughout a array of PPs.

\begin{figure}[ptbh]
 \begin{centering}
        \includegraphics[scale = 0.35,trim = 10 330 160 195, angle = -0,clip]{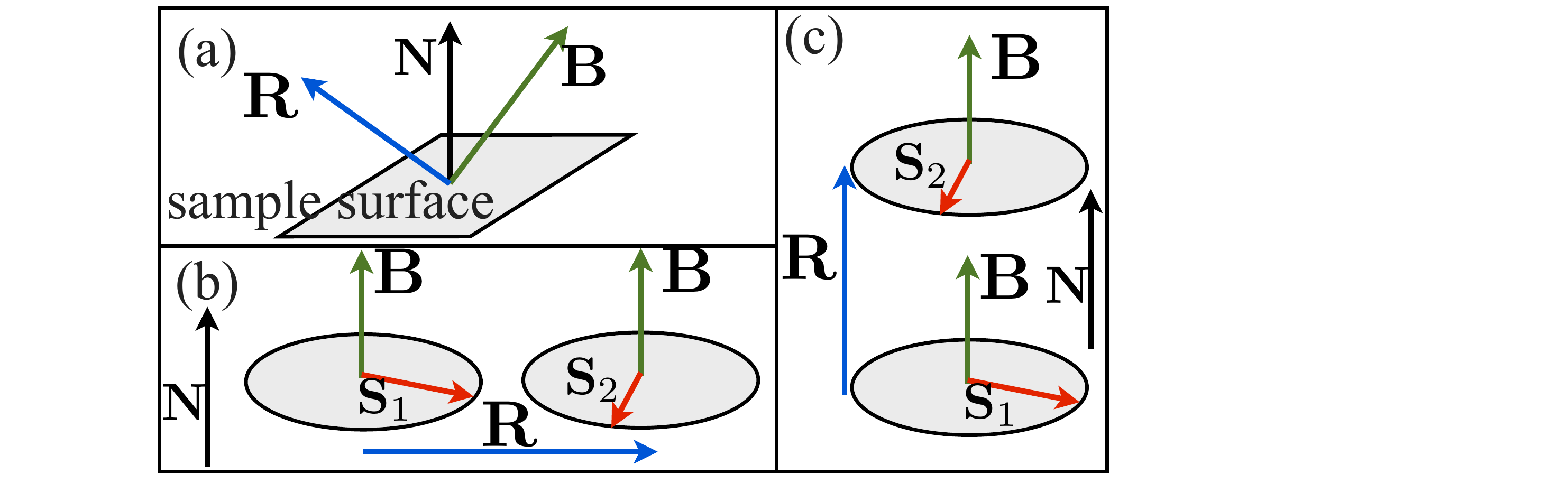}
        \caption[]
{(Color online) (a) There are three vectors of importance: $\bold{N}$, the vector normal to the sample surface, the applied field $\bold{B}$, and displacement vector $\bold{R}$. (b) $\bold{R}$ is lateral and (c) $\bold{R}$ is parallel to direction of current ($\bold{N}$).  
In all, the two spins are denoted by their projections on the plane orthogonal to the applied field.
}\label{fig:dipolarToon0}
        \end{centering}
\end{figure}

Spin carrying polarons in the vicinity of one another experience each other's spin magnetic moment. 
For simplicity we assume point dipoles.
The energy of the interaction depends on the orientation of the spins. 
This magnetic interaction is known as the dipolar interaction with the Hamiltonian
\begin{equation}
\mathscr{H}_{dip} = D (\bm{S}_1 \cdot \bm{S}_2 - 3 (\bm{S}_1 \cdot \hat{R})(\bm{S}_2 \cdot \hat{R}))
\end{equation}
where $D = \mu_0 g^2 \mu_B^2/4 \pi \hbar R^3$. $\bm{R}$ is the vector displacement between the two polaron spins.
If $\hat{R}$  was uniformly distributed between PPs, one does not expect any OMAR angular dependence;\cite{ODea2005} however there is reason to believe that there is an anisotropy in the formation of PPs such that $\hat{R}$ is restricted to a narrower range of directions;
specifically $\hat{R}$ may align with the sample normal as shown in Figure \ref{fig:dipolarToon0}(c).\cite{Wagemans2011}
This is a sensible expectation since an electric field should direct carrier flow normal to the sample thereby increasing the probability of parallel (to electric field) spin-blocking (Figure \ref{fig:dipolarToon0}(c)) as opposed to lateral (Figure \ref{fig:dipolarToon0}(b)) spin-blocking (in plane of sample).
Following previous researchers\cite{Efimova2008, Wagemans2010, Wagemans2011} we make the assumption that the exchange and dipolar couplings are constants for each PP.
A more general description of inhomogeneous spin-spin couplings is an  avenue for future work.

\section{Percolation Theory of OMAR}\label{section:model}

A key development in electronic transport theory in disordered media was the random resistor network model proposed by Miller and Abrahams over a half century ago.\cite{Miller1960} In the simplest form where only spatial disorder is considered, the resistance between any two sites, $i$ and $j$, is given by $R_{ij} = R_0 e^{2 r_{ij}/\ell}$ where $r_{ij}$ is their separation and $\ell$ is the localization length of a carrier at a site. 
Miller and Abrahams were unsuccessful in determining the bulk resistance in their model; in fact it was not for over a decade that the resistance was calculated correctly using a relatively new mathematical theory known as percolation theory.\cite{Ambegaokar1971, Pollak1972, Shklovskii1984} The arguments of percolation theory dictate that bulk resistance is governed by a critical resistance (distance) $R_c$ ($r_c$) which is the smallest resistance (separation) that permits an infinitely large chain of sites connected by resistors (separations) less than the critical one.
The threshold length is constrained by the bonding criterion: 
\begin{equation}\label{eq:bonding1}
 \int_0^{r_c} 4 \pi N r^2 dr = B_c,
\end{equation}
 where $N$ is the density of sites and $B_c$ tells how many bonds each site must connect to on average to be included in the percolating network; $B_c \approx 2.7$ in three dimensions.\cite{Shklovskii1984}
Conduction within $r$-percolation has been observed in organic semiconductors in the regime of large inter-site separations and high temperatures where the influence of energy disorder is minimized.\cite{Gill1974, Rubel2004} These conditions are assumed throughout this article. Despite the fact that transport in most organic semiconductors may be determined by energy disorder alone,\cite{Cottaar2011} our results elsewhere\cite{Harmon2012a, Harmon2012b} and herein share many qualitative elements with OMAR experiments.
This correspondence suggests that at least some aspects of OMAR may be independent of disorder type.
 
\begin{figure}[ptbh]
 \begin{centering}
        \includegraphics[scale = 0.35,trim = 60 260 150 150, angle = -0,clip]{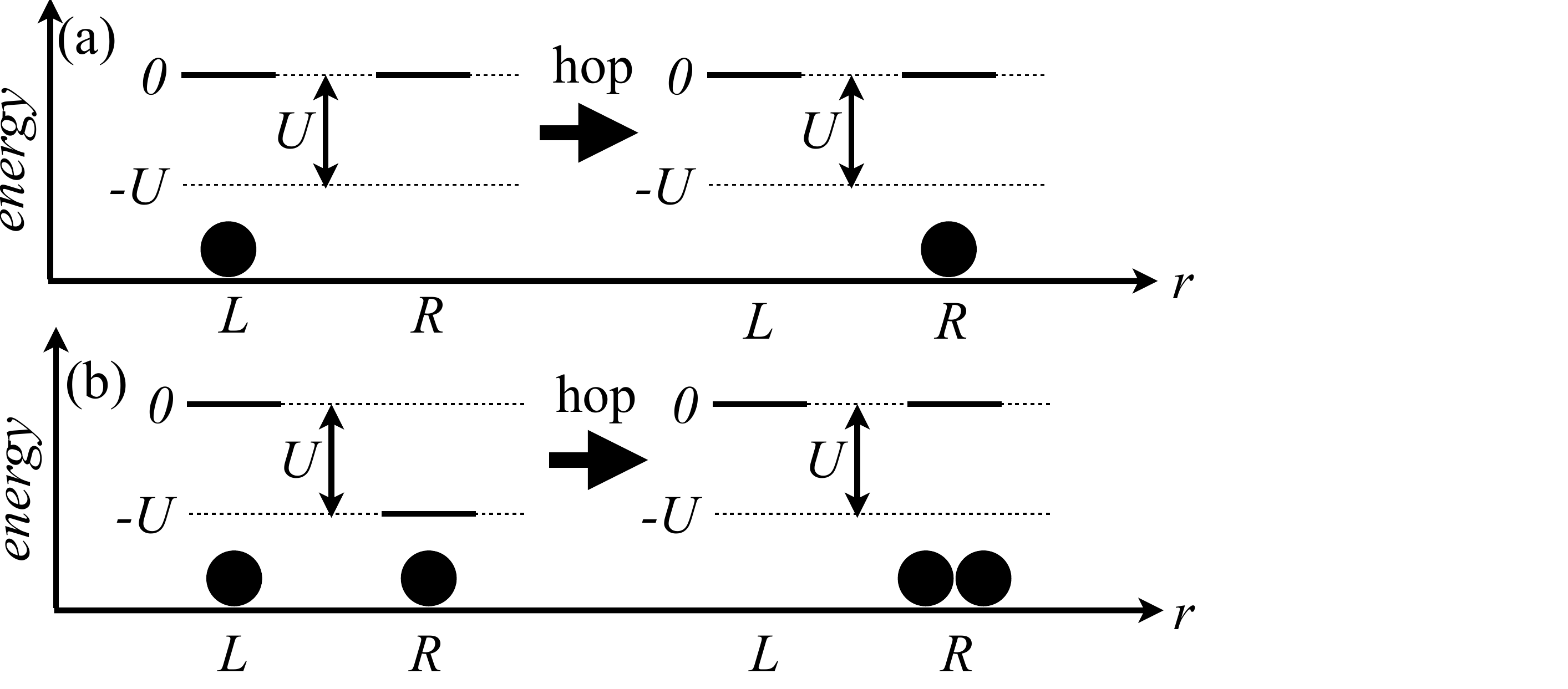}
        \caption[]
{Energy description of hopping between sites. Energy of a polaron/bipolaron is shown as a thick line. The dotted lines mark specific energy levels zero and $-U$. Solid circles represent actual carriers in space.
(a) Polaron at L hops to unoccupied site R. Both sites are at equivalent energies.
(b) Polaron at L hops to occupied site R which was at energy $-U$. The Coulomb interaction energy of the bipolaron lifts the \emph{total} energy back to zero.
}\label{fig:hubbard}
        \end{centering}
\end{figure}

For completeness, the spin-dependent percolation theory\cite{Harmon2012a,Harmon2012b} is summarized here. In this theory, as also in many other OMAR theories, spin affects hopping transport through the Pauli exclusion principle; occupation of two polarons at a single site (a bipolaron) is forbidden in the triplet state (T), but permitted in the singlet state (S) at an energy expense $U$ - the Coulomb interaction energy. For $r$-percolation alone, including $U>0$ provides a charge-blocking effect that diminishes spin-flip effects because bipolaron formation costs energy $U$ regardless of the relative spin orientation. Since in reality energy disorder entails a spread in site energies (comparable to $U$) the charge-blocking effect is unphysical since there are frequently bipolaron energy states resonant with the polaron energy. 
This is shown schematically in Figure \ref{fig:hubbard}.
So to avoid any unphysical charge-blocking effects, we make assumption that $U=0$. 
A more sophisticated approach would examine the problem with energy disorder and/or variable range hopping\cite{Osaka1979},  but is beyond the scope of this current work.

A polaron with arbitrary spin is restricted from hopping to another polaron if the PP's spin state is T but may hop to another polaron (forming a bipolaron) if the  result is an S state just as if to an unoccupied site. For a polaron looking to perform a hop, the respective densities of these three types of sites are $N_{T}$, $N_{S}$, and $N_{0}$. 
Since the spins of forming PPs are random, spin statistics mandates that $N_S = \frac{1}{4} N_1$ and $N_T = \frac{3}{4} N_1$, where $N_1$ is the concentration of injected polarons. 
$N_1$ is small compared to the total concentration of sites so that a polaron encountering a bipolaron is unlikely.

A PP that exists in the T state cannot form a bipolaron; this restriction effectively reduces the concentration of sites to $N_{eff}' = N  - N_T$ since only these sites are accessible to a hopping polaron. When spin transitions are absent, the bonding criterion of Eq. (\ref{eq:bonding1}) becomes $N\rightarrow N_{eff}'$.
One expects the resistance in the sample to increase since accessible sites are now typically further away from one another.
Whether a PP combines into a bipolaron or disassociates as the polarons hop away from one another, the dynamics are integrally dependent on spin transitions since bipolaron formation is blocked for T states but unblocked for S states.
The effects of spin transitions are now considered.
Spin-blocked paths can be opened if the PP's spin function changes from T to S.
A given polaron hopping to another site occupied by a polaron is successful 1/4 of the time.
The other 3/4 of the time the hopping attempt is rejected due to the PP being in the T state.
At this point two things may occur:
\begin{itemize}
\item[] \emph{Bipolaron formation:} The polaron waits until the next hop. If a spin transition (T-S) has occurred then the hop is successful.
\item[] \emph{Disassociation:} If the spin transition has not taken place and a more distant site is available, the polaron will hop to this further site.
\end{itemize}
\noindent There is a trade-off between the two processes; if sites are well-packed or the hopping attempt frequency is very rapid, the polaron is likely to disassociate.
If the opposite is true (dilute and slow hopping frequency), the polaron is prone to wait until the spin transition occurs.\cite{Harmon2012a, Harmon2012b}
Therefore spin transitions adjust the number of blocked sites from $N_T$ to $(1 - p_{T\rightarrow S})N_T$ where  $p_{T\rightarrow S}$ is the probability for the blockade to cease when the next hop is attempted. 
This probability may depend on time, spatial separation, applied field, and other system parameters depending on what interaction initiates the spin transitions. 
With the inclusion of spin transitions the bonding criterion becomes
\begin{equation}\label{eq:bondingCriterion1}
\int_0^{r_c}4 \pi  N_{eff} r^2 dr = B_c 
\end{equation}
with $N_{eff} = N  - N_T + p_{T\rightarrow S} N_T$. 
After defining the MR to be MR$= \langle [R_c(B) - R_c(0)]/R_c(0) \rangle$, we use Eq. (\ref{eq:bondingCriterion1})
to write\cite{Harmon2012b}
\begin{equation}\label{eq:fullMR}
\textrm{MR}
\approx \frac{2}{3} \frac{1}{y_{c_1}^2}\eta \int_0^{y_{c_1}}  y^2  \big\langle p_{S}(B)  - p_{S}(0) \big\rangle dy,
\end{equation}
where we have made the substitution $p_{T\rightarrow S} = \frac{1}{3} [1 - p_{S}]$ in terms of the easier to evaluate quantity $p_{S\rightarrow S} \equiv  p_S$ and $\eta = N_T/N$.\cite{Werner1977} 
This equation is valid only in the dilute limit ($N_1 \ll N$) but this situation is believed to be the physically realized one.
We have used a dimensionless length $y = r/\ell$.
The renormalized critical length is
$y_{c_1} = y_{c_0} (1-N_T/N)^{-1/3}$ where $y_{c_0} = (3B_c/4 \pi \ell^3 N)^{1/3}$ is the critical length in the spinless problem.
$y_{c_1}$ is nearly $y_{c_0}$ because of the assumption of dilute polarons. 
In any case, we use the threshold as an independent variable and refer to it as simply $y_c$  hereon.
Angular brackets denote an appropriate averaging which in our case is over hyperfine fields.

The task of calculating the MR now reduces to finding the probabilities for singlets at time, $t$, that were initiated in the singlet state $p_{S}(B)$.
PPs do not experience a hop at exactly $\tau_h$; some hop earlier and some later and the probability that a hop has not occurred at time, $t$, is $\exp(-t/\tau_h)$.\cite{Schulten1978}
Therefore $p_{S}(B) = \frac{1}{\tau_h} \int_0^{\infty} \rho_{S}(t) e^{-t/\tau_h} dt$ where $\rho$ is the two-spin density matrix.
Spin-spin interactions affect the MR by altering the spin transition probabilities.

We now consider the modifications required due to the new total Hamiltonian $\mathscr{H}$ from Sec.~\ref{section:semiclassical} to obtain results for the MR. We must determine, from the total Hamiltonian, the singlet occupation probability at a time $t$ given an initial singlet state ($\rho_S$).
This was previously accomplished by using\cite{Harmon2012b}
\begin{equation}\label{eq:oldDensityMatrix}
\rho_{S} = |\langle S| \exp(-i \mathscr{H} t) |S \rangle|^2.
\end{equation}
Direct computation could follow after decomposing the exponentiated operator using the Euler formula.
The key property of the Hamiltonian that allowed this process was that it was Zeeman-like.
This is no longer the case with the spin-spin Hamiltonians now involving exchange interactions and dipolar terms.
In fact no progress can be made using Ref. \onlinecite{Harmon2012b}'s technique since
\begin{equation}\label{eq:nonEuler}
e^{i (\mathcal{H}_0 + \mathcal{H}_{exc} + \mathcal{H}_{dip})t} \neq e^{i\mathcal{H}_0 t} e^{i(\mathcal{H}_{exc} + \mathcal{H}_{dip})t}
\end{equation}
due to a lack of commutativity between Hamiltonians.
If we wish to obtain analytic results, we are forced then to assume small spin-spin interactions and use perturbation theory.

\section{Effect of spin-spin interactions on MR in the fast hopping limit}\label{section:spinspin}

To ascertain analytic expressions in this section, we assume $\hat{R}||\hat{N}$ and that these vectors subtend an angle $\alpha$ from the applied field which we take to be in the $\hat{z}$ direction.
Then for small $a_{eff}$, $J$, and $D$, but arbitrary $\omega_0$, the spin density matrix can be solved analytically using perturbation theory in the interaction representation.
The non-interacting case ($J =0$, $D = 0$) has been examined in the past\cite{Haberkorn1977, Harmon2012b},
and we also solve it  below  as an  instructional example.
To the best of our knowledge, the interacting case has not been dealt with in this manner prior to the treatment here.

In the interaction representation, the following operators are defined:
\begin{equation}
\mathscr{H}_{1}^{*}(t) = e^{i \mathscr{H}_Z t} \mathscr{H}_{1} e^ {-i \mathscr{H}_Z t},
\end{equation}
where $\mathscr{H}_1 = \mathscr{H}_{hf} + \mathscr{H}_{exc} + \mathscr{H}_{dip}$ is the perturbing Hamiltonian.
Similarly, the new spin density matrix is
\begin{equation}
\rho^{*}(t) = e^{i \mathscr{H}_Z t} \rho(t) e^ {-i \mathscr{H}_Z t};
\end{equation}
initially it can be shown that $\rho^{*}(0) = \rho(0)$.

To simplify we consider the following: all terms odd in the hyperfine interaction vanish when the hyperfine interaction is assumed to be isotropic; additionally, it can be shown that the \emph{first, second, and third} order contributions from the spin-spin interactions do not affect the singlet occupation probability $\rho_S(t)$.
The influence of exchange and dipolar interactions must be examined to fourth order perturbation theory. The hyperfine interaction also contains a fourth order contribution.

Several terms do not contribute and are neglected when we write out the perturbed density matrix:\cite{Slichter1996}
\begin{widetext}
\begin{eqnarray}\label{eq:4thorder}
&&\rho^{*}(t) = \rho(0) - \int_0^t \int_0^{t'}[[\rho(0), \mathscr{H}_{hf}^{*}(t')], \mathscr{H}_{hf}^{*}(t'')]dt'' dt' +{}\\
&&{} \int_0^t \int_0^{t'} \int_0^{t''} \int_0^{t'''} \Big[\Big[\big[\big[\rho(0), \mathscr{H}_{hf}^{*}(t')], \mathscr{H}_{hf}^{*}(t'') + \mathcal{H}_{exc}^{*}(t'') + \mathcal{H}_{dip}^{*}(t'')],\mathscr{H}_{hf}^{*}(t''') +   \mathcal{H}_{exc}^{*}(t''') + \mathcal{H}_{dip}^{*}(t''')\Big],  \mathscr{H}_{hf}^{*}(t'''')\Big]dt'''' dt''' dt'' dt'.\nonumber
\end{eqnarray}
\end{widetext}

\subsection{Analytic solution in absence of exchange or dipolar interactions}

We begin by considering the solution in the absence of exchange or dipolar interactions. The hyperfine Hamiltonian in the interaction representation is $\mathscr{H}_{hf}^{*}(t)  =  U(\mathscr{H}_{{hf}_1}+\mathscr{H}_{{hf}_2})U^{\dagger}$ with $U = e^{i \mathscr{H}_Z t} = [\cos\frac{\omega_0 t}{2} + 2 i S_1^z \sin\frac{\omega_0t}{2}] [\cos\frac{\omega_0 t}{2} + 2 i S_2^z \sin\frac{\omega_0 t}{2}]$ since $\mathscr{H}_{{hf}_1}$ commutes with $\mathscr{H}_{{hf}_2}$.
We write $\rho(0) = P_S$ where $P_S$ is the singlet projection operator:
 \begin{eqnarray}\label{eq:singletProjection}
P_S = 
\left( {\begin{array}{cccc}
 1& 0 &0 & 0 \\
0 & 0 &0 & 0 \\
 0 & 0 &0 & 0\\
 0 & 0 &0 & 0
 \end{array} } \right)
\end{eqnarray}
in the $S$, $T_0$, $T_+$, $T_-$ basis.

The singlet part of the density matrix $\langle S | \rho^{*}(t)|S \rangle$ is  $\rho_S(t)$.
Initializing in the singlet state requires that $\rho_S(0) = 1$.
The first order correction vanishes.
After averaging over the hyperfine fields for the ensemble of two carriers, including the second order term, Eq. (\ref{eq:4thorder}), yields
\begin{equation}\label{eq:arbField}
\langle \rho_S  \rangle = 1 - \frac{1}{8} a_{eff}^2 t^2 \bigg[1+2 \frac{\sin^2 \omega_0 t/2}{(\omega_0 t/2)^2}\bigg]
\end{equation}
in agreement with the quantum mechanical calculation.\cite{Haberkorn1977, Salikhov1984} 

To find the MR, we substitute the singlet probability, Eq. (\ref{eq:arbField}), into Eq. (\ref{eq:fullMR}). 
After integrating over the exponential distribution of hopping times,
\begin{equation}\label{eq:MR1}
\textrm{MR} =\frac{1}{3}\eta
\frac{1}{ y_{c}^2 r^2} \int_0^{y_{c}}  \frac{ \omega_0 ^2 }{\omega_0 ^2 + 1/\tau_h^2} y^2 dy,
\end{equation}
where $r = (\tau_h a_{eff})^{-1}$ and $\tau_h^{-1} = v_0 \exp{(-2 y)}$.
The integral in Eq. (\ref{eq:MR1}) can be computed when the hopping rate has an exponential dependence on the hopping distance $\tau_h^{-1} = v_0 \exp(-2 y)$.
The result is cumbersome but can be considerably simplified under the usual assumption that $y_c \gg 1$ to
\begin{equation}\label{eq:osaka}
 \textrm{MR}  =
\frac{1}{12}  \eta \frac{1}{r^2_c}\big[1 - \frac{1}{\omega_0 ^2 \tau^2_c}\ln(1+\omega_0^2 \tau^2_c) \big],
\end{equation}
where $r_c = (\tau_c a_{eff})^{-1}$ and $\tau_c^{-1} = v_0 \exp{(-2 y_c)}$.
At saturating fields, $\textrm{MR}_{sat}/\eta =  1/(12 r_c^{2})$.

\subsection{Analytic solution with exchange and dipolar interactions}

We now consider the addition of exchange and dipolar interactions. By using commutator algebra, the fourth order hyperfine interaction can be separated from the spin-spin interactions which enter in quadratically with a quadratic hyperfine term.
This implies that the spin-spin interactions would have no effect on the singlet probability if not for coexisting with the hyperfine interaction.
This is understood by recognizing that the spin-spin Hamiltonians do not couple the singlet state with triplet states.

The mathematics simplifies when no magnetic field is applied.
The second order correction can be obtained from Eq. (\ref{eq:arbField}) in the appropriate limit.
The time-averaged singlet probability up to fourth order obtained from Eq. (\ref{eq:4thorder}) is
\begin{equation}\label{eq:zeroField}
\langle p_S(B = 0) \rangle = 1 - \frac{3}{4} \frac{1}{r^2} + \frac{33}{16} \frac{1}{r^4}  +  \frac{3}{8} \frac{1}{r^4} (d^2 + 8 j^2),
\end{equation}
where the second term is the second order contribution and $d = D/a_{eff}$, and $j = J/a_{eff}$.

The effect of exchange and dipolar interactions is to lift the degeneracy of singlet and triplet states.
In strongly applied magnetic fields, the $T_+$ and $T_-$ states are far in energy from the $S$ and $T_0$ states due to the large Zeeman splitting.
Consequently, only transitions between  $S$ and $T_0$ states occur; the problem simplifies considerably since a smaller subspace of the original Hamiltonian warrants examination. 
The 2$\times$2 Hamiltonian matrix,
 \begin{eqnarray}\label{eq:subspace}
&&\mathscr{H}_{hf} + \mathscr{H}_{exc} + \mathscr{H}_{dip} ={}\nonumber\\ 
&&\left( {\begin{array}{cc}
 -J & (\omega_1^z - \omega_2^z)/2  \\
(\omega_1^z - \omega_2^z)/2  &J+ {D}{}(1+3 \cos 2\alpha)/4   \\
 \end{array} } \right),
\end{eqnarray}
is readily diagnolizable;
the exchange splitting magnitude of $S$ and $T_0$ is $\omega_{S-T_0} = a_{eff} |d(1-3 \cos^2 \alpha) - 4 j|/2$.
The time averaged singlet-to-singlet probability is
\begin{equation}\label{eq:infiniteField}
\langle p_S(B\rightarrow \infty) \rangle = 1 - \frac{1}{4} \frac{1}{r^2}  + \frac{3}{8} \frac{1}{r^4} + \frac{1}{4} \frac{1}{r^4} \frac{\omega_{S-T_0}^2}{a_{eff}^2}.
\end{equation}
To compare with Ref. \onlinecite{Wagemans2011}, it is useful to define a quantity to measure the effects of angle variations:
\begin{equation}
\Theta (B, \alpha) \equiv  \frac{\langle \Delta p(B,\alpha) \rangle}{\langle \Delta p(B,0) \rangle} \equiv \frac{\langle p_{T\rightarrow S}(B, \alpha)\rangle - \langle p_{T\rightarrow S}(0) \rangle}{ \langle p_{T\rightarrow S}(B, 0) \rangle - \langle p_{T\rightarrow S}(0) \rangle}.
\end{equation}
\begin{figure}[ptbh]
 \begin{centering}
        \includegraphics[scale = 0.8,trim = 0 2 0 0, angle = -0,clip]{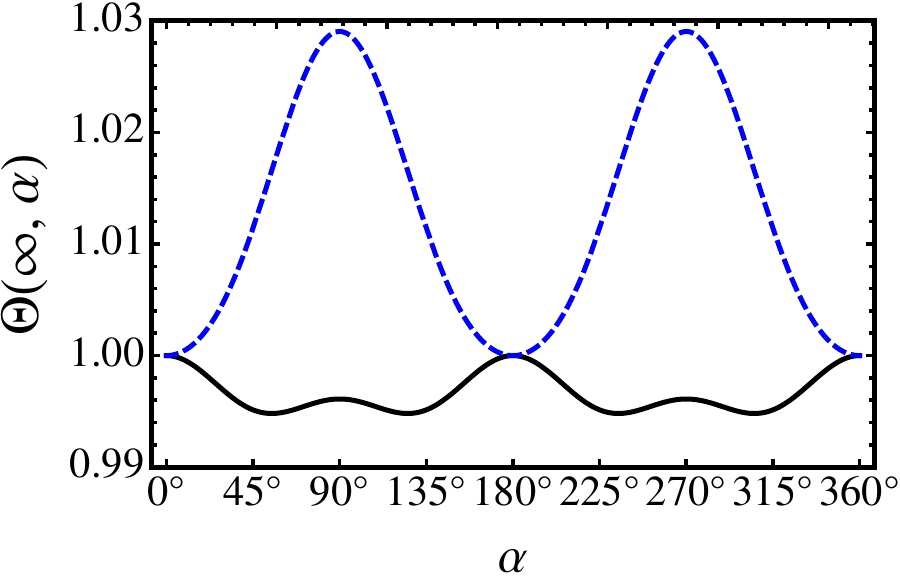}
        \caption[]
{(Color online) Angle dependence of singlet probability. Solid black: $j = 0$; dashed blue: $j = -1 d$. For both, used $d = 1$ and $r = 5$. For ease of comparison with Ref. \onlinecite{Wagemans2011}, $\tau_h$ has been assumed constant.}\label{fig:angleNorm}
        \end{centering}
\end{figure}
The function $\Theta$ is plotted for two values of the exchange constant in Figure \ref{fig:angleNorm}.
The number of maxima is contingent on the exchange coupling; the simulations of Wagemans \emph{et al.} found that for $J \geq -\frac{1}{2}D$ four maxima occur.\cite{Wagemans2011}
By examining our analytic expression for $\Theta$, the conditions of $J$ that alter the number of stationary points can be obtained from $d\Theta /d\alpha = 0$.  We find the analytic condition for the existence of the four maxima,
\begin{equation}
-\frac{1}{2} D \leq J \leq \frac{1}{4} D.\label{4maxima}
\end{equation}
The simulations of Ref. \onlinecite{Wagemans2011} identified the existence of the lower bound, but here we report the full range where four maxima would be observed.
The fourth order expression in a general field is available in closed form in the Appendix.
We connect the occupation probabilities to the transport problem as described in Section \ref{section:model}.
For any magnetic field, the MR can be expressed in closed form, although not in a concise manner; it can be found in the Supplementary Information due to its length.

At saturating fields,
\begin{equation}
\langle p_S(\infty, \alpha) - p_S(0, \alpha) \rangle =  \frac{1}{2 r^2} - \frac{27}{16r^4} - \frac{3 d^2+ 24 j^2- \frac{2\omega_{S-T_0}^2}{a_{eff}^2}}{8 r^4}.
\end{equation}
The full result is burdensome but
without much error we can write down a simplified expression for the saturated MR:
\begin{eqnarray}\label{eq:4thOrder}
\frac{\textrm{MR}_{sat}}{\eta} &=& \frac{\textrm{MR}_{sat}(d = 0, j = 0)}{\eta}(1 - \frac{1}{2y_c}) {}\nonumber\\
&&{} - \frac{1}{192}\frac{1}{r_c^4}(27+6 d^2 + 48 j^2 - 4 \frac{\omega_{S-T_0}^2}{a_{eff}^2}),
\end{eqnarray}
where $\textrm{MR}_{sat}(d = 0, j = 0)/\eta = 1/(12 r_c^2)$. 
The angular dependence completely lies in the term $\omega_{S-T_0}^2$ causing the saturated MR to have different number of maxima depending on $J$ and $D$ just as with the case of $\Theta$.
The angular dependence of the saturated MR can be examined analytically:
\begin{eqnarray}
\frac{\textrm{MR}_{sat}(90^{\circ})}{\eta} - \frac{\textrm{MR}_{sat}(0^{\circ})}{\eta} &=& \frac{1}{48}\frac{1}{r_c^4}\frac{( \omega_{S-T_0}^2(90^{\circ}) - \omega_{S-T_0}^2(0^{\circ}))}{a_{eff}^2}{}\nonumber\\
&=&{}  -\frac{1}{24}\frac{1}{r_c^4}d(\frac{3}{4}d+j),
\end{eqnarray}
where $J$ can be seen to control whether the difference in responses is positive or negative.
Additionally, the angular response evaporates as hopping rate increases which leads us to the inference that spin-spin interactions and their concomitant angle-of-field dependence are unlikely to be observed experimentally in the fast hopping limit.

\section{Effect of spin-spin interactions on MR for arbitrary hopping rates}\label{section:general}

\subsection{Calculation of the singlet probability}

As mentioned earlier, we are unable to use the analysis of Ref. \onlinecite{Harmon2012b} when spin-spin interactions are included.
In the fast hopping we avoided the problem by using perturbation theory.
However for arbitrary hopping times, we must use a different approach to finding the singlet probability.
We begin again with Eq. (\ref{eq:oldDensityMatrix}); by inserting two complete sets of eigenstates of the full Hamiltonian
the singlet portion of the density matrix is
\begin{eqnarray}\label{}
\rho_{S} &=& \sum_{m, m'}\langle S| \exp(-i \mathscr{H} t)|m\rangle \langle m |S \rangle \langle S |  \exp(i \mathscr{H} t)|m' \rangle \langle m' | S \rangle {}  \nonumber\\
{} &=& \sum_{m, m'}e^ {i (\omega_{m'} - \omega_{m}) t} \langle S|m\rangle \langle m |S \rangle \langle S  | m' \rangle \langle m' | S\rangle{} \nonumber\\
&=&{} \sum_{m, m'}e^{i (\omega_{m'} - \omega_{m}) t} | P_S^{mm'} |^2 \nonumber ,
\end{eqnarray}
where $P_S =  |S \rangle \langle S  |$ is the S projection operator and its matrix representation is Eq.~(\ref{eq:singletProjection}).
Since the desired quantity is $p_S = \frac{1}{\tau_h} \int_0^{\infty} \rho_{S}(t) e^{-t/\tau_h} dt$, we first compute the time integral which can be readily accomplished:
\begin{equation}\label{eq:fullPs}
p_{S} =
 \sum_{m, m'}| P_S^{mm'} |^2 \frac{ 1/\tau_h^2}{ (\omega_{m'} - \omega_{m})^2 + 1/\tau_h^2}.
\end{equation}
This result is completely general for any Hamiltonian.
This expression entails that only states with some singlet component contibute to $p_S$.
In this form, the spatial integral of Eq. (\ref{eq:fullMR}) can be calculated yielding the result:
\begin{eqnarray}\label{eq:fullMR2}
&&\int_0^{y_c} y^2 dy ~\langle p_S(B, y\rangle= \Big\langle \sum_{m, m'}| P_S^{mm'} |^2 \times \nonumber\\
&&{} \Bigg[\frac{y_c^3}{3} - \frac{y_c^2}{4}\ln(1+\frac{(\omega_{m'} - \omega_{m})^2}{v_0^2 \exp(-4 y_c)}) - \frac{y_c}{8} \textrm{Li}_2(-\frac{(\omega_{m'} - \omega_{m})^2}{v_0^2 \exp(-4 y_c)}) -\nonumber \\
&&{}\frac{1}{32}\textrm{Li}_3(-\frac{(\omega_{m'} - \omega_{m})^2}{v_0^2}) +
\frac{1}{32}\textrm{Li}_3(-\frac{(\omega_{m'} - \omega_{m})^2}{v_0^2 \exp(-4 y_c)}) \Bigg]\Big\rangle.
\end{eqnarray}
$\textrm{Li}_n(x)$ is the polylogarithm function of order $n$.\cite{Lewin1981}
The importance of the polylogarithm terms dwindles as the hopping rate increases.
In the case of slow hopping, such that $v_0 \rightarrow 0$, only terms in Eq. (\ref{eq:fullPs}) such that $m' =m$ contribute to the MR which reduces the right-hand side of Eq. (\ref{eq:fullMR2}) to $\frac{y_c^3}{3}\big\langle \sum_{m}| P_S^{mm} |^2  \big\rangle$.
While the expression appears simple, the eigenvectors and eigenvalues of our Hamiltonian are complex enough that analytic results are not possible. 
\begin{figure}[ptbh]
 \begin{centering}
        \includegraphics[scale = 0.9,trim = 0 0 0 5, angle = -0,clip]{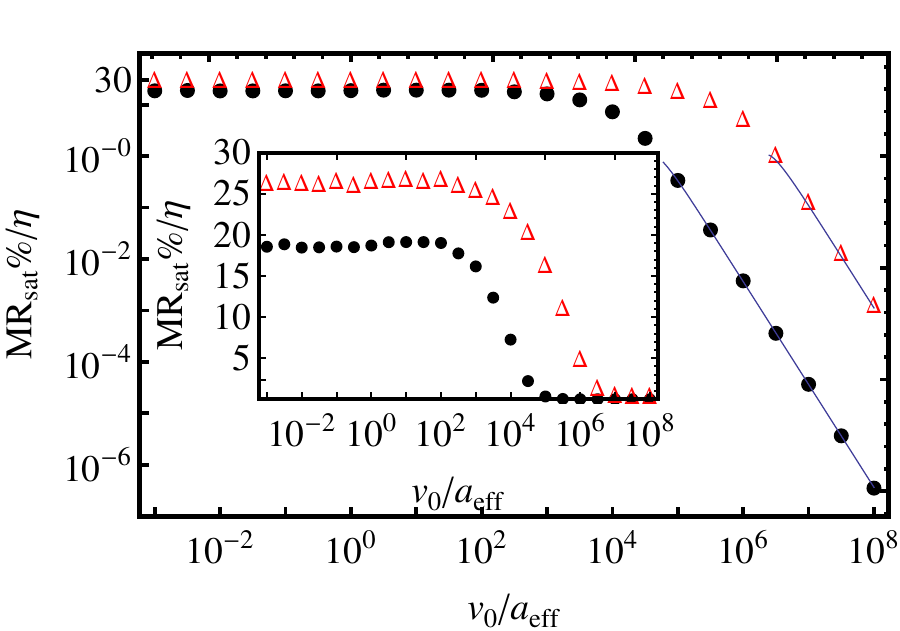}
        \caption[]
{(Color online) MR$_{sat}$ as a function of hopping rate with no spin-spin interactions. Black solid circles: $y_c = 5$; red triangles: $y_c = 7$. Solid lines are analytic result, Eq. \ref{eq:4thOrder}, from 4th order perturbation theory. Inset is same as main except on log-linear axes. Averaging required $10^3$ configurations of the hyperfine fields. Results agree with alternate method of Ref. \onlinecite{Harmon2012b}.}\label{fig:mrSatnon}
        \end{centering}
\end{figure}

We first test this formalism against the results obtained in Ref. \onlinecite{Harmon2012b}.
The saturated MR, with spin-spin interactions absent, is plotted in Figure \ref{fig:mrSatnon} and agrees with the prior formalism used in Ref. \onlinecite{Harmon2012b}.

For the remainder of this article, the $\hat{z}$ direction is fixed to be parallel to $\hat{N}$.
We restrict $\hat{R}$ to lie parallel to $\hat{N}$ except where we allow $\hat{R}$ to vary over a range of angles.
We choose to isolate the effects of the exchange and dipolar interactions by examining them independently in this and the next subsection.
In the final subsection~\ref{section:exchangeAndDipolar} both interactions are analyzed in tandem.

\subsection{Exchange interaction}

\begin{figure}[ptbh]
 \begin{centering}
        \includegraphics[scale = 0.33,trim = 20 265 30 75, angle = -0,clip]{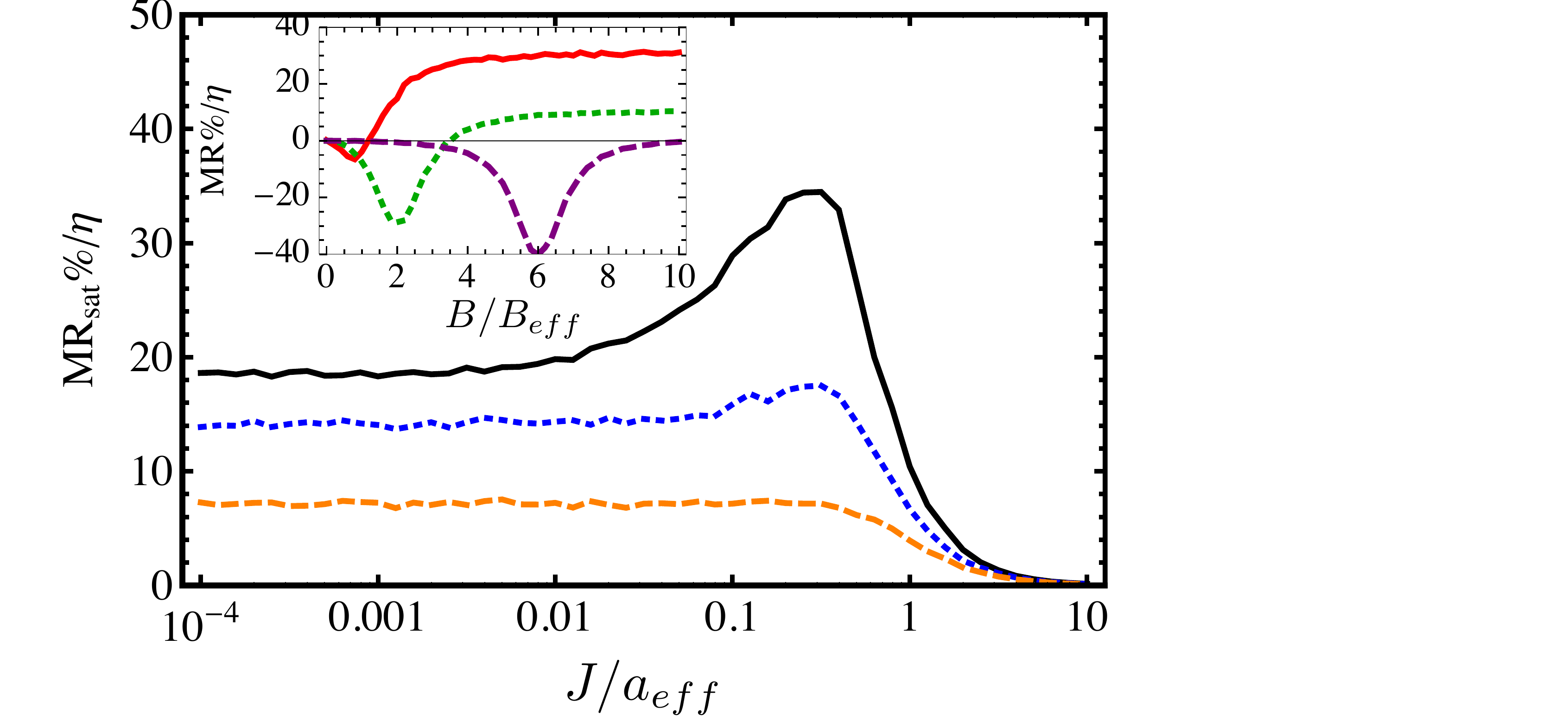}
        \caption[]
{(Color online) Saturated MR versus exchange constant for different values of the hopping rate; solid black: $v_0/a_{eff} = 0$, dotted blue: $v_0/a_{eff} = 2\times 10^3$, dashed orange: $v_0/a_{eff} = 10^4$. For all, $y_c = 5$.  Inset: MR in the slow hopping limit for several different values of the exchange constant:  solid red: $J=  0.4$, dotted green: $J=  1$, dashed purple: $J=  3$.  For all, $y_c = 5$.}\label{fig:slowJ}
        \end{centering}
\end{figure}

Figure \ref{fig:slowJ} plots the saturated MR versus exchange constant for several different hopping rates.
As expected from the non-interacting picture, the overall MR decreases with hopping rate.
When we concentrate on the \emph{zero hopping limit} (solid black line in Figure \ref{fig:slowJ}), it is noteworthy that at small and large exchange couplings the results agree well with the Figure \ref{fig:mrSatnon} for small and large hopping rates.

We interpret this as the exchange interaction acting much like hopping; instead of the carrier hopping site to site, the spin of the carriers swaps between occupied sites. This heuristic understanding is further supported by  the calculation of $\langle p_S(B\rightarrow \infty) \rangle $, wherein we find exchange can be accounted for exactly by the substitution $1/\tau_h \rightarrow 1/\tau_h+ 2 J$.
 If $J/a_{eff}$ is small, the hyperfine fields fully mix spin before the spins swap and the case is no different than for slow hopping and no exchange. 
If $J/a_{eff}$ is large, spins swap back and forth faster than hyperfine fields can mix which leads to a motional narrowing effect much like what occurred when the hopping was very rapid (but hopping is effectively zero here). 
This effect has been used in doped inorganic semiconductors to explain certain long spin lifetimes due to motional narrowing by exchange interactions.\cite{Dzhioev2002, Kavokin2008}
A non-zero hopping rate suppresses the overall saturated MR as discussed in Ref. \onlinecite{Harmon2012b}.

The large MR enhancement prior to $J/a_{eff} = 1$ in Figure \ref{fig:slowJ} is explainable in the following way:
as $J$ increases from zero, both $p_S(B\rightarrow \infty)$ and $p_S(B=0)$ increase sharply when $J \sim a_{eff}$ since the amount of spin mixing will be significantly reduced when the exchange interaction surpasses the hyperfine interaction.
The enhancement is due to the fact that both functions do not increase in the same way; $p_S(B=0)$ increases more slowly with $J$ than $p_S(B\rightarrow \infty)$. 
As $J$ increases far past $a_{eff}$, the zero and infinite field probabilities become nearer in value (unity) since, as in the case of faster hopping, the hyperfine fields produce less spin mixing in both cases.
The reason for the lag in $p_S(B = 0)$ is readily apparent when considering the pertinent energy levels and Eq. (\ref{eq:fullPs}).
At zero field, all four electronic energy levels are roughly degenerate when $J\sim a_{eff}$. In a large field, two energy levels are already well separated and lead to negligible spin mixing.
Therefore it takes a larger $J$ to suppress spin mixing in zero field than it does in large field.

Eq. \ref{eq:fullPs} shows how energy level placement affects spin mixing;
energy levels well separated from the $S$ level do not factor into spin mixing (and $p_S$ increases toward unity).
In the absence of exchange, increasing the applied field separated out the $T_+$ and $T_-$ states from the $T_0$ and $S$ states which is why MR$>$0 (since $p_S(B) > p_S(0)$).
Inclusion of exchange allows for an USFE: when $\omega_0 \sim 2 |J|$, the $T_+$ or $T_-$ levels are near the $S$ or $T_0$ levels, which increases the amount of spin mixing such that $p_S(B) < p_S(0)$ (which leads to MR$<$0).
MR is shown for three different exchange constants in Figure \ref{fig:slowJ}'s inset.
For finite yet small $J$  (black solid line), the USFE develops and becomes more pronounced as $J$ further increases (dotted green and dashed purple lines).
The exchange interaction may serve as the explanation of the USFE measured in unipolar organic devices.\cite{Nguyen2010b}
Increasing the hopping rate degrades the large negative MR produced by the large exchange interaction (not shown).

\subsection{Dipolar interaction}

\begin{figure}[ptbh]
 \begin{centering}
        \includegraphics[scale = 0.93,trim = 0 0 0 0, angle = -0,clip]{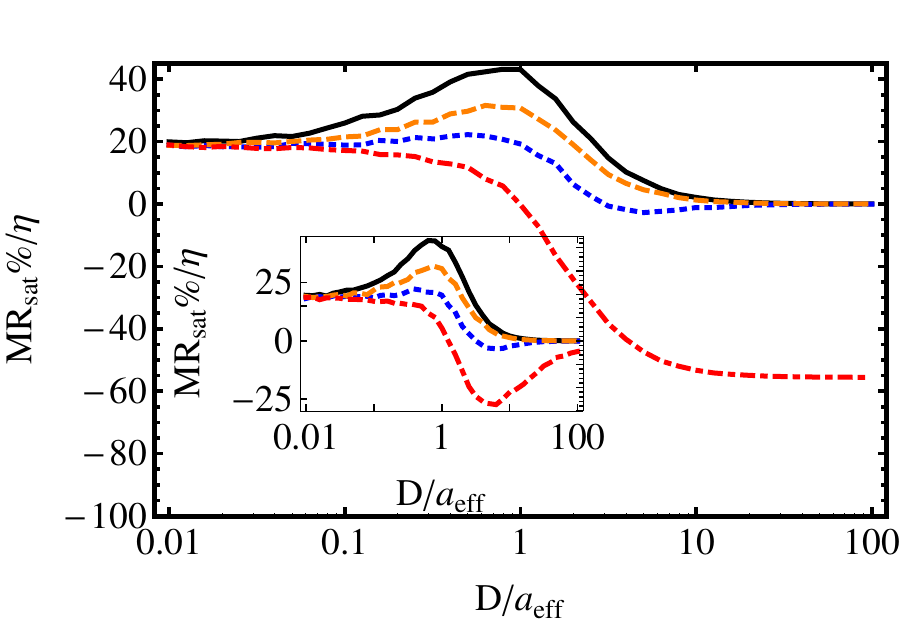}
        \caption[]
{(Color online) Saturated MR plotted at $\alpha = 0^{\circ}$ (solid black), $\alpha = 45^{\circ}$ (dotted blue), $\alpha = \frac{1}{2} \cos^{-1} (-\frac{1}{3}) \approx 55^{\circ}$ (dash-dotted red), and $\alpha = 90^{\circ}$ (dashed orange). Hopping rate and exchange are zero. Non-zero hopping rate reduces overall MR$_{sat}$. Inset: same as main except polaron pair displacement vector $\bold{R}$ is not fixed but is averaged over an angular range within 10$^{\circ}$ of the $\hat{z}$.}\label{fig:dipolar1}
        \end{centering}
\end{figure}

In this section $J$ is set to zero and the dipolar interaction constant $D$ is turned on.
We focus here solely on the slow hopping limit.
Figure \ref{fig:dipolar1} depicts the dependence of saturated MR versus $D/a_{eff}$ on the angle-of-field, $\alpha$. 
The difference is dramatic for $D/a_{eff} \sim 1$. 
At angles around $\frac{1}{2} \cos^{-1} (-\frac{1}{3}) \approx 55^{\circ}$ the saturated MR goes strongly negative.
At the angle $\frac{1}{2} \cos^{-1} (-\frac{1}{3})$ the dipolar splitting vanishes in the high field region; a unique situation is created in which the zero-field splitting of the energy levels is larger than the high field splitting between the $T_0$ and $S$ levels. This ensures more hyperfine mixing and less resistance at high fields than at zero field which lead to negative MR. 
Such large angular dependence has not been observed.
This may be explainable by the inset of Fig. \ref{fig:dipolar1} where it can be seen that the special angle loses its uniqueness when the dipolar coupling is allowed to vary within 10$^{\circ}$ of $\hat{N}$. Therefore it is unlikely that the special angle could be observed in experiments since the displacement vectors between PPs are not fixed but vary over a range of angles.

The angular dependence is plotted as dashed red curves in Figure \ref{fig:anglePlot} from $\alpha = 0^{\circ}$ to $180^{\circ}$.
The concavity of the blue (solid) curves is set by the sign of the exchange interaction; hence, if the angle dependence is due to spin-spin interactions, angle measurements of OMAR could indicate the sign of the exchange interaction if the direction of the dipolar anisotropy is established.
In the two insets, the strong angular dependence due to the special angle, $\frac{1}{2} \cos^{-1} (-\frac{1}{3})$, is seen to diminish as the PP orientations are made to be less anisotropic. In the limit that the PPs orientations are completely isotropic (not shown), the angular dependence vanishes as expected.

\subsection{Exchange and dipolar interactions}\label{section:exchangeAndDipolar}

\begin{figure}[ptbh]
 \begin{centering}
        \includegraphics[scale = 0.93,trim = 0 0 0 0, angle = -0,clip]{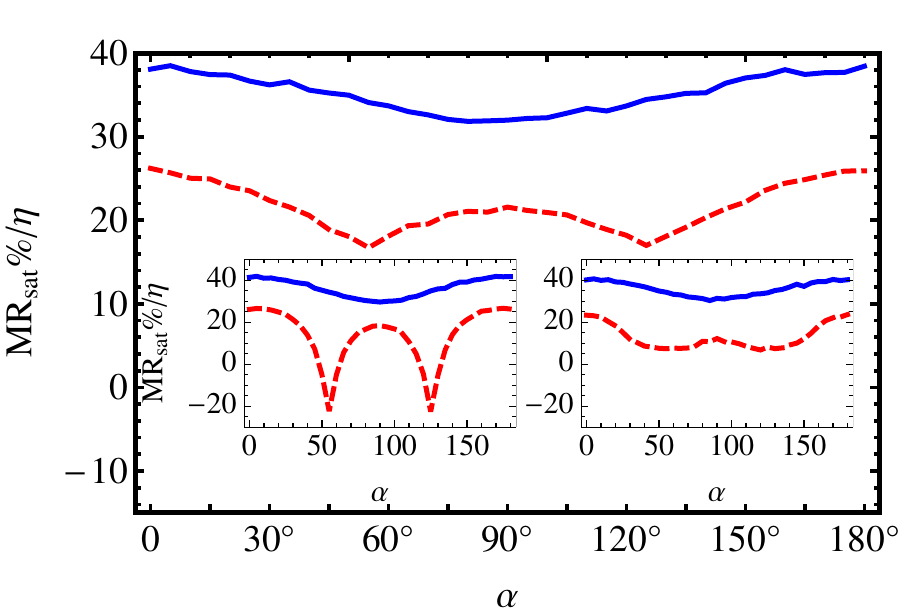}
        \caption[]
{(Color online) Main: solid blue: $d = 0.1$, $j = 0.2$; dashed red: $d = 0.1$; $j = 0$. Inset left: solid blue: $d = 0.2$, $j = 0.2$; dashed red: $d = 2$, $j = 0$. Inset right: same as left inset except polaron pair displacement has an angular spread around $\hat{z}$ of 35$^{\circ}$. $4\times 10^{3}$ hyperfine configurations were used in the averaging procedure.}\label{fig:anglePlot}
        \end{centering}
\end{figure}

In the regime opposite to the one considered in Section \ref{section:spinspin} (i.e. slow hopping/large spin-spin coupling), we numerically determine that the same conditions on $J$ and $D$ found in Section \ref{section:spinspin}, Eq.~(\ref{4maxima}), are still valid and produce either two or four maxima in the angle (0$^{\circ}$ to 360$^{\circ}$) versus MR$_{sat}$ graph as shown in the main portion of Figure \ref{fig:anglePlot}.
These maxima continue to occur since the high field dipolar splitting varies with angle which in turn varies the amount of $S-T_0$ mixing; however the modulation now depends on $J$ and $D$: $\alpha =\frac{1}{2} \cos^{-1} (-\frac{1}{3}-\frac{8 J}{3D})$ and is not pinned near $55^{\circ}$ as was the case without exchange.

An important conclusion to draw is that \emph{if} the angle dependence of MR is due to dipolar coupling, the exchange interaction must be present and non-negligible since zero exchange yields the wrong number of maxima.
Neither can the exchange term be much greater than the dipolar term since it will suppress MR in that case (not shown).
So the assertion that the exchange and dipolar couplings affecting MR are comparable seems necessary despite the drastic difference in their spatial dependencies.\cite{Efimova2008, Schellekens2010}
As already stated, a dipolar interaction initiated angle dependence is contingent on there being a PP orientation anisotropy.

\section{Conclusions}

Recent experiments have measured an angle-of-field dependent magnetoresistance. 
One possibility for this effect is an anisotropic dipolar coupling between hopping polarons.
To evaluate the role of dipolar coupling in these systems, we have calculated magnetoresistance in positionally disordered unipolar organic systems.
In accord with other authors,\cite{Wagemans2011} the magnetoresistance response to the dipolar interaction must be examined in conjunction with exchange coupling to agree with experiments.
Sorting out the cause of the magnetoresistance dependence on angle-of-field requires experimentation in more organic systems that may allow for variations in the spin-spin couplings. 
Observation of four maxima would be unequivocal evidence for the dipolar interaction being the source of magnetoresistance variation due to angle.
Another possibility for the angle dependence is the existence of an anisotropic hyperfine interaction.
The theory described here can be modified to account for anisotropic hyperfine distributions and is the subject of future work.

The theory outlined above is for unipolar organic devices.
Many experiments are believed to be in a bipolar current regime.
We previously argued\cite{Harmon2012b} that our theory is still qualitatively useful in describing bipolar MR with the relation MR$_{unipolar}\sim - \lambda$ MR$_{dipolar}$ where $\lambda$ is some proportionality factor.
This is identical to saying MR$_{unipolar}\sim \lambda$ MC$_{dipolar}$.

In summary, we have calculated the effects of spin-spin interactions on magnetoresistance in positionally disordered organic semiconductors. 
The magnetoresistance shows an angle-of-field dependence which agrees with experiments for certain ratios of the exchange and dipolar parameters. 
Future experimentation is needed to decide whether the angle dependence is due to the dipolar coupling or some other anisotropic mechanism.
In addition to the utility of these calculations to the fledgling field of organic spintronics,
we remark that these calculations of singlet and triplet probabilities including both the exchange and dipolar interactions may find application in the study of magnetic field effects on chemical reactions and more specifically their manifestation in avian navigation mechanisms.\cite{Timmel1998, Ritz2009}

\section{Acknowledgements}

This work was supported by an ARO MURI. We thank P. A. Bobbert for answering several questions regarding Ref. \onlinecite{Wagemans2011}.

\appendix
\section{$4^{\text{th}}$ Order Singlet Probability}
The fourth order singlet probability is composed of a hyperfine only term ($p_1^{(4)}$) and a mixed term ($p_2^{(4)}$) of hyperfine and spin-spin couplings:
$p_S^{(4)} = p_1^{(4)} + p_2^{(4)}$ where
\begin{equation}
\langle p_1^{(4)} \rangle  = 
\frac{24 h^8+90 h^6 r^2+145 h^4 r^4+136 h^2 r^6+33 r^8}{16 r^4 \left(h^2+r^2\right)^3 \left(4
   h^2+r^2\right)},
\end{equation}
and
\begin{widetext}
\begin{eqnarray}
\langle p_2^{(4)} \rangle = &&
\frac{1}{128 r^4
   \left(h^2+r^2\right)^3}
\Big[h^6 \left(11 d^2+32 d j+128 j^2\right)+3 h^4 \left(11 d^2+32 d j+128 j^2\right) r^2+48 h^2
   \left(d^2+4 d j-8 j^2\right) r^4+{}\nonumber\\
   &&{}48 \left(d^2+8 j^2\right) r^6+12 d h^2 \left(h^4 (d+8 j)+3 h^2 (d+8
   j) r^2+48 j r^4\right) \cos (2 \alpha )+9 d^2 h^4 \left(h^2+3 r^2\right) \cos (4 \alpha )\Big],
\end{eqnarray}
\end{widetext}
with $h = \omega_0/a_{eff}$.
MR can be calculated from these equations but due to their length, they are located in the Supplementary Information.

\end{document}